\documentclass[aps,prl,twocolumn,superscriptaddress]{revtex4-1}
\usepackage{graphicx}
\usepackage{caption}
\usepackage{subcaption}

\usepackage[font=small,labelfont=bf]{caption}
\usepackage[margin=1in]{geometry}

\usepackage{amsmath}
\usepackage{hyperref}
\usepackage{floatrow}
\usepackage{siunitx}
\usepackage {ulem}

\usepackage{ragged2e}
\usepackage{dcolumn}
\usepackage{bm}

\begin{document}

\preprint{APS/123-QED}

\title{Dynamics of current-induced switching in the quantum anomalous Hall effect}
\author{Alina Rupp}
\author{Daniel Rosenbach}
\author{Torsten Röper}
\author{Dominik Hoborka}
\author{Alexey A. Taskin}
\author{Yoichi Ando}
\author{Erwann Bocquillon}%
 \email{bocquillon@ph2.uni-koeln.de}
 \affiliation{II. Physikalisches Institut, Universit\"at zu K\"oln, Z\"ulpicher Str. 77, D-50937 K\"oln, Germany}
\date{\today}

\begin{abstract}
Ferromagnetic topological insulators in the quantum anomalous Hall (QAH) regime host chiral, dissipationless edge states whose propagation direction is determined by the internal magnetization. Under suitable conditions, a strong electrical bias can induce magnetization reversal, and thus flip the propagation direction. In this work, we perform time-resolved measurements to investigate the switching dynamics. Our results reveal characteristics consistent with a disordered magnetic landscape and demonstrate that the reversal process is thermally activated, driven by Joule heating during the current pulse. The understanding of the magnetization dynamics in QAH systems opens pathways for local, controlled manipulation of chiral edge states via thermal effects.
\end{abstract}

\maketitle


\section{Introduction}

The quantum anomalous Hall (QAH) effect in ferromagnetic topological insulators arises from a single chiral topological edge state, whose propagation direction is set by the out-of-plane magnetization of the sample. This results in vanishing longitudinal resistance $R_{xx}\simeq 0$ and a quantized Hall resistance $R_{xy}= \pm \frac{h}{e^2}$ without the need for an external magnetic field \cite{chang_experimental_2013, chang_high-precision_2015, Yu2010}. Owing to these properties, QAH materials are interesting playgrounds for quantum transport in edge plasmons \cite{martinez_edge_2023, roeper2024} or for induced topological superconductivity \cite{uday2023}. They offer a promising platform for quantum metrology \cite{Bestwick-2015, gotz2018, Fox-2018, fijalkowski2024}, for the development of non-reciprocal microwave components \cite{viola_hall_2014, mahoney_zero-field_2017} or flying Majorana states \cite{beenakker_deterministic_2019}.

Crucially, understanding, controlling, and reversibly switching the magnetization in these ferromagnetic systems would enable directional control of the edge state, facilitating path reconfiguration \cite{yasuda2007} and the creation of interfaces at magnetic domain walls \cite{rosen2017}. In this context, current-induced magnetization switching has been reported in two separate studies, though under remarkably different conditions -- one compatible with spin-orbit torque mechanisms \cite{Yuan2024} and the other not \cite{Zimmermann2024}.

In this work, we employ time-resolved excitation pulses to reveal the dynamics of current-induced magnetization switching in thin films of V-doped (Bi,Sb)$_2$Te$_3$ (V-BST) in the quantum anomalous Hall regime. Thus, we access the switching time $t_{\rm sw}$ necessary for the magnetization to reverse under excitation, and clarify the underlying switching mechanism. In our devices, the observed behavior is inconsistent with spin-transfer torques. Instead, by modeling heat transport within our films, we demonstrate that magnetization reversal results from thermally-activated switching, originating from Joule heating. Specifically, in the presence of a small perpendicular magnetic field opposing the magnetization of the sample, current pulses sufficiently raise the temperature to exponentially decrease the switching time $t_{\rm sw}$, and trigger the reversal of multiple independent domains, therefore progressively leading to a full reversal of the magnetization.

\begin{figure}
    \centering
    \includegraphics[width=\linewidth]{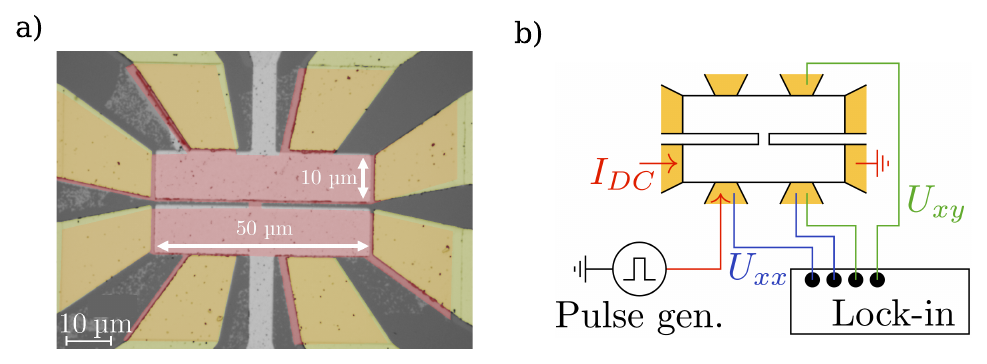}
    \caption{\justifying\textbf{Experimental setup:} a) Microscope image of Sample A. The mesa is colored in red. Parts of it are covered by a gate, not functional in this device. The ohmic contacts are colored in light yellow. b) Schematic of the measurement setup. A pulse generator ({\it Rigol DG5102}) delivers short pulses of duration $\tau$ between \SI{25}{\nano\second} and \SI{100}{\micro\second} applied to the sample via a coaxial line. Lock-in amplifiers measure the longitudinal and transverse (Hall) resistance of the sample in between the pulsed excitations.}
    \label{fig:device_setup}
\end{figure}

\section{Experimental setup}

\begin{figure*}[htp!]
    \centering
    \includegraphics[width=\linewidth]{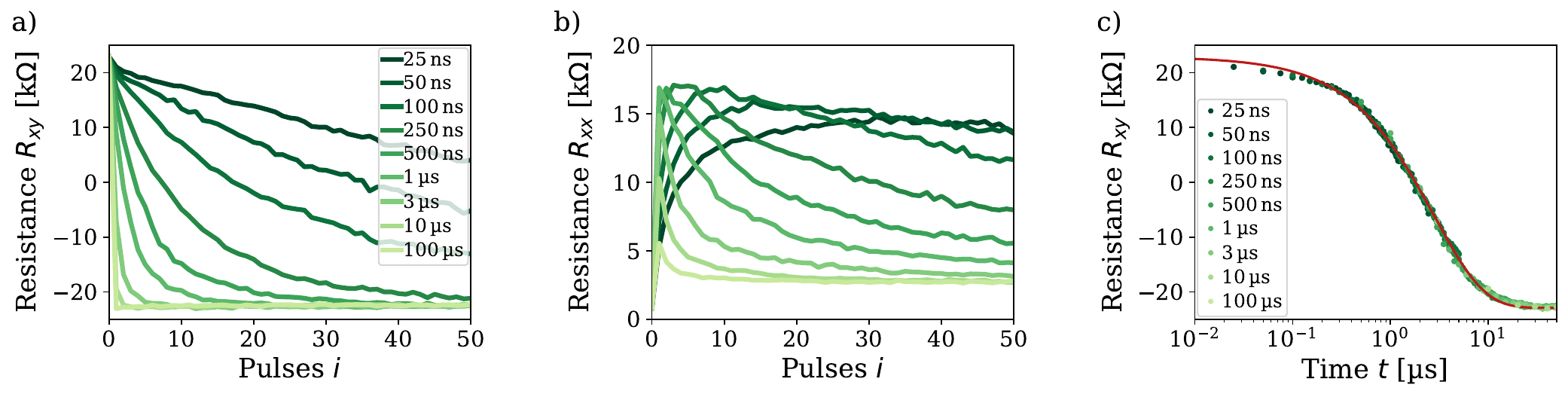}
    \caption{\justifying\textbf{Dynamics of the resistances:} a) (resp. b)) Evolution of the resistance $R_{xy}$ (resp. $R_{xx}$)  measured after an increasing number $i$ of pulses, for an amplitude $V=\SI{4}{\volt}$ for different pulse widths $\tau$ (shown in different shades of green, legend visible in c)), measured on Sample A. c) Dynamics of $R_{xy}$ as a function of the time $t=N\tau$ spent under excitation, showing a very good overlap between the curves obtained for different pulse widths $\tau$. A fit of the dynamics of $R_{xy}(t)$ with a stretched exponential function is plotted as a red solid line, for $t_{\rm sw}=\SI{2.8}{\micro\second}$ and $\beta = 0.8$.}
    \label{fig:PW_dep}
\end{figure*}

In this work, several devices were measured, all fabricated using \SI{8}{\nano\meter} thick films of V-BST grown by molecular beam epitaxy, capped with a \SI{4}{\nano\meter} layer of $\rm Al_2O_3$. In this article, we focus on this article in two main devices, Sample A and B. A microscope image of Sample A is shown in Fig.~\ref{fig:device_setup}a. All transport measurements were performed \SI{20}{\milli\kelvin}, at the base temperature of the refrigerator. Ohmic contacts are used for measuring longitudinal and transverse Hall resistances (resp. $R_{xx}$ and $R_{xy}$) in different configurations as allowed by the sample geometry. We use standard lock-in techniques at low frequency (\SI{7}{\hertz}) and low amplitude (\SI{1}{\nano\ampere}) (see Fig.\ref{fig:device_setup}b). Each device exhibits a Hall resistance within approximately \SI{10}{\percent} of the quantized value $R_{xy}=h/e^2$, and longitudinal resistance values that range between $R_{xx} \simeq\SI{300}{\ohm}$ and \SI{600}{\ohm}. In addition, one of the ohmic contacts is connected to a high-frequency coaxial line with low attenuation (maximum \SI{-18}{\deci\bel}). Via this coaxial line, we send pulses of large amplitude up to $V\simeq\SI{10}{\volt}$ (at the generator output) and of temporal widths $\tau$ ranging between $\SI{25}{\nano\second}$ and $\SI{100}{\micro\second}$. The corresponding currents, in excess of \SI{10}{\micro\ampere}, are similar to those used in Ref.\cite{Zimmermann2024}.

\section{Experimental observations on current-induced switching}

\paragraph{Current-induced switching monitored from the magneto-resistance --}

We start by describing the measurement sequences and their results. In such a sequence, the sample is first initialized at a large external field ($\mu_0 H_{\rm init}=\SI{2}{\tesla}$) to a fully magnetized quantum anomalous Hall state, in which $R_{xy}\simeq R_K$ and $R_{xx}<\SI{1}{\kilo\ohm}$, before setting the magnetic field to a value $H$ such that $-H_c<H<0$ (where $\mu_0 H_c\simeq \SI{1}{\tesla}$ is the coercive field), opposing the initial magnetization $M_{\rm init}>0$. We repeat pulses of amplitude $V$ and duration $\tau$ and measure the evolution of $R_{xx}, R_{xy}$ in between each two pulses to observe the progressive reversal of the magnetization. The sequence ends after $N$ pulses. The measurements of $R_{xy}$ and $R_{xx}$ are shown in Figs. \ref{fig:PW_dep}a and \ref{fig:PW_dep}b for various pulse durations. As the number of applied pulses $i\leq N$ increases, the Hall resistance $R_{xy}$ is observed to progressively decrease, until it saturates at the value $R_{xy}\simeq-R_K$ expected for a film fully magnetized in the opposite direction. In parallel, $R_{xx}$ shows a strong increase for small $i$, before decreasing to $R_{xx}\simeq\SI{3}{\kilo\ohm}$ approaching the value observed in the initial state. This indicates that, during the reversal of the magnetization, the film transits through the magnetically unpolarized state in which quantum anomalous Hall edge states wrap around the magnetic domains and dissipation occurs. As in this example, the longitudinal resistance $R_{xx}$ is often observed to converge towards final values higher than the initial one. We discuss possible origins for this difference in the Supplementary Material \cite{supplement}.

\paragraph{Magnetization dynamics --}
We now discuss the dynamics of the reversal of the Hall resistance, and indirectly of the magnetization. We first rescale each curve by defining the time coordinate $t$ indicating the total cumulated time under excitation conditions, via the discrete points $i\tau$, for $i\leq N$. We observe that, regardless of the chosen pulse width $\tau$, all the measured curves $R_{xy}(t)$ overlay (see Fig.\ref{fig:PW_dep}c). Importantly, it shows that the resistances and magnetization only depend on the total time $t$ spent under the voltage drive, but do not evolve in between. We have additionally tested multiple pulsing configurations (varying the intervals between pulses for example) to confirm this fact (see \cite{supplement}). It proves that such a stroboscopic measurement of $R_{xy}$ after each pulse provides a reliable determination of the dynamics of the magneto-resistance, and thus indirectly of the magnetization. We base our analysis on this robust observation.

Following this observation, we investigate the dynamics of the current-induced switching in different conditions. We adapt the duration of the pulses $\tau$ to the experimental observations to access both short timescales ($t<\SI{100}{\nano\second}$) as well as longer ones ($t\gg\SI{10}{\milli\second}$) within reasonable measurement times \footnote{Alternatively, when the RF line is equipped with bias-tee limiting the bandwidth, we also used trains of multiple short pulses.}. We find that the time evolution of $R_{xy}(t)$ is, over more than 3 decades, in very good agreement with the stretched exponential form:
\begin{equation}
    R_{xy}(t)=R_0\Big(1-2\,\exp\big(-(\frac{t}{t_{\rm sw}})^\beta\big)\Big)
    \label{eq:strexp}
\end{equation}
An example fit is shown in Fig.\ref{fig:PW_dep}c, with here $R_0=\SI{23}{\kilo\ohm} \simeq R_K$, $\beta=0.8$ and $t_{\rm sw}=\SI{2.8}{\micro\second}$. To lowest order, one can assume that $R_{xy}\propto M$ (following the law of the classical anomalous Hall effect). The stretched exponential behavior then suggests a magnetically strongly disordered system \cite{Phillips1996,Xi2008,Lucas2022}, in which multiple small domains are reversed independently, with a broad distribution of switching times. Stretching parameters as low as $\beta\simeq0.4$ have been observed in some cases. This observation is compatible with earlier works which evidenced the strong magnetic disorder in the parent compound Cr-doped BST \cite{Lee2015}.

\paragraph{Scaling laws --}

\begin{figure}
    \centering
    \includegraphics[width=\linewidth]{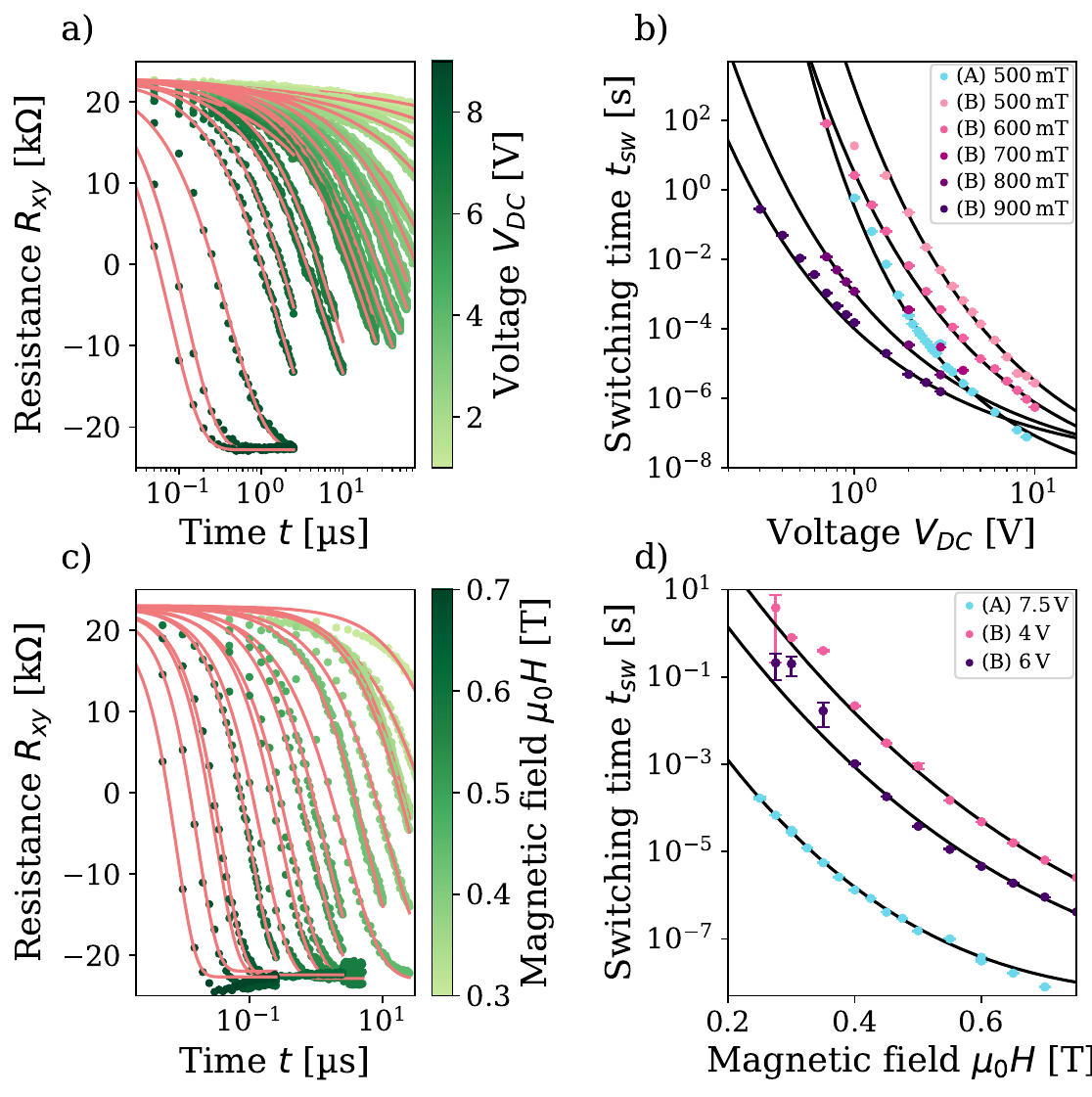}
    \caption{\justifying\textbf{Voltage and magnetic field dependencies:} a) Measurements of $R_{xy}$ as function of $t$, for various pulse amplitudes $V$ (shown in different shades of green), taken on Sample A, showing strong changes in the dynamics. The red lines show fits of the data sets with Eq.(\ref{eq:strexp}). b) Extracted switching times $t_{\rm sw}$ as function of pulse amplitude $V$, for Sample A ($\mu_0 |H|=\SI{500}{\milli\tesla}$) and Sample B ($\mu_0 |H|=500, 600, 700, 800, \SI{900}{\milli\tesla}$). The black solid lines show fits to Arrhenius activation laws for each data set. c) Measurements of $R_{xy}$ as function of $t$, for various magnetic fields $H$, taken on Sample A, similarly showing strong changes in the dynamics. d) Extracted switching times $t_{\rm sw}$ as function of magnetic field $H$, for Sample A ($V=\SI{7.5}{\volt}$) and Sample B ($V=$ 4 and \SI{6}{\volt}), and fits to Arrhenius activation laws as black solid lines.}
    \label{fig:scaling}
\end{figure}

Using this fit function, we can extract the switching time $t_{\rm sw}$ for varying magnetic fields $H$ and amplitudes $V$. First, we observe as illustrated in Fig.\ref{fig:scaling}a that $t_{\rm sw}$ decreases very strongly with $V$: $R_{xy}$ evolves and switches much faster for high voltages. The fits to the stretched exponential decays show that the switching time $t_{\rm sw}$ varies by approximately 9 orders of magnitude over the measurable decade of voltage $V\simeq 0.1 - \SI{1}{\volt}$, ranging from \SI{0.1}{\micro\second} to \SI{100}{\second}. Higher voltages may damage the sample, while low voltages result in extremely slow dynamics ($t_{\rm sw}\gg\SI{100}{\second}$), which is inaccessible in a reasonable time or without long pulses and significant heating of the entire refrigerator. For reproducible and comparative studies, we have added waiting times to ensure that the mixing chamber temperature remains constant below \SI{15}{\milli\kelvin} regardless of the number or duration of the pulses. The extracted switching times $t_{\rm sw}$ for Sample A as well as for Sample B are reported in Fig.\ref{fig:scaling}b, and will be further analyzed in the next section. We note that the stretched exponential fits do not work as well for Sample B but the value of $t_{\rm sw}$ remains correctly extracted and its behavior remains unchanged (see also \cite{supplement}). 

We now turn to measurements of the switching time $t_{\rm sw}$ as a function of the external magnetic field $H$ as presented in Fig.\ref{fig:scaling}c and \ref{fig:scaling}d, for a fixed pulse amplitude $V$. The spread of $t_{\rm sw}$ is also very spectacular. We capture changes of $t_{\rm sw}$ over 4 orders of magnitude for magnetic fields opposed to the magnetization with $\mu_0 |H|$ between 0.3 and \SI{0.7}{\tesla}, ranging from approximately \SI{0.05}{\micro\second} to \SI{10}{\second}. We analyze these variations in the next section.

\section{Heat-driven mechanism for switching}

\paragraph{Role of spin-transfer torques --} Before presenting a toy model for thermally activated current-induced switching, we comment on the role of spin-transfer torques. Our observations are fully incompatible with reversal under the action of spin-transfer torques exerted by the electron flow, as reported in \cite{Yuan2024}. First, experiments performed under in-plane magnetic fields did not exhibit any magnetic switching. Second, the orientation of the pulse current, with respect to the external magnetic field or the magnetization, does not have an influence (see \cite{supplement}). For example, it is not possible to switch the magnetization reversibly by a change in the current direction.
In the next section, we show that our observations are on the contrary compatible with a heat-driven mechanism, in which the electron temperature strongly increases under the application of the voltage pulse, thereby dramatically decreasing the switching time of individual domains, and finally leading to a full reversal of the magnetization.

\paragraph{\label{sec:theory} Model for heat-induced reversal --}
  
\begin{figure}
    \centering
    \includegraphics[width=\linewidth]{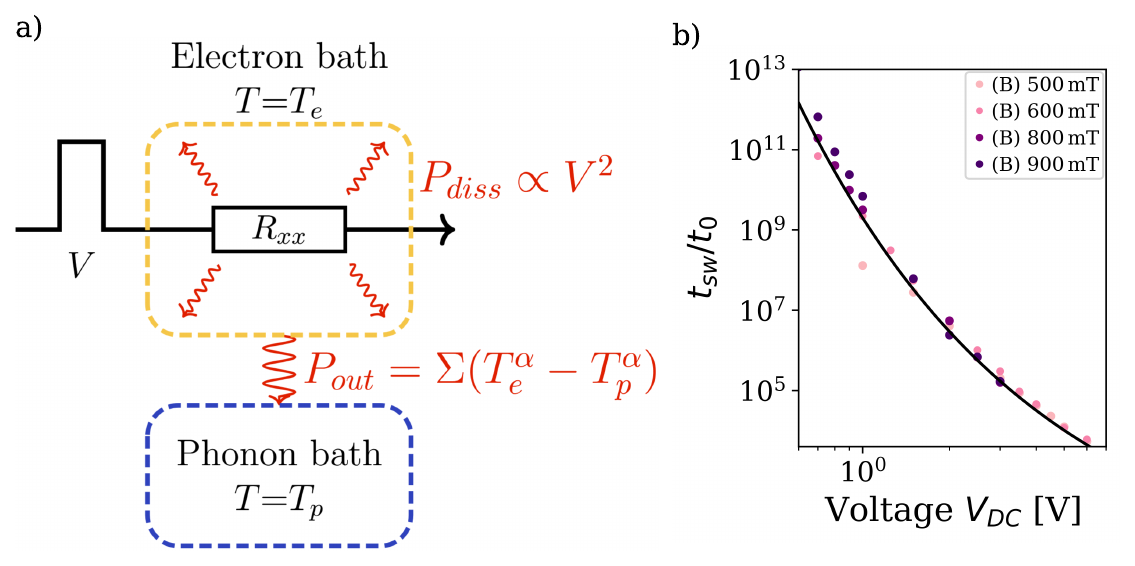}
    \caption{\justifying\textbf{Model for thermally activated switching:} a) The dissipated heat $P_{\rm diss}$ resulting from the application of the pulse increases the electronic temperature $T_e$. The electrons are cooled via phonon emission with the power $P_{\rm out}$, depending both on $T_e$ and on the lattice temperature $T_p$. b) Normalized switching time $t_{\rm sw}/t_0$, subsequently rescaled as described in the main text, showing the agreement between the different data sets and the validity of the Arrhenius law over 8 decades.}
    \label{fig:model}
\end{figure}

In this section, we introduce a simple model for thermally activated current-induced switching, depicted in Fig.\ref{fig:model}a. The pulse of large amplitude $V$ drives large currents in the sample which strongly exceed the breakdown threshold of the QAH effect \cite{lippertz-2022,roeper2025}. Thus, the resistance of the sample $R_{xx}$ increases and the heat is generated by the Joule effect, with a power $P_{\rm diss}\propto V^2$. We assume that the electron bath immediately equilibrates to a temperature $T_e$ while the lattice remains at equilibrium at the base temperature of the refrigerator $T_p\ll T_e$. Under this sudden elevation of temperature, the magnetization locally switches to align with the external magnetic field $H$. This scenario converts to the following predictions. In semiconducting devices at low temperatures, cooling through electron-phonon scattering is known to follow the typical power law $P_{\rm out}=\Sigma\big(T_e^\alpha-T_p^\alpha)$ with $\Sigma$ a constant and the exponent $\alpha$ in the range $3$ to $5$ \cite{jezouin2013, lebreton2022, Ferguson2025,roeper2025}. Equating $P_{\rm diss}=P_{\rm out}$ at thermal equilibrium thus yields $T_e\propto V^{2/\alpha}$ in the limit of high temperatures $T_e\gg T_p$.
At this temperature $T_e$, the typical switching time is given by an Arrhenius law of the form $t_{\rm sw}(H,T_e)=t_0 \exp\left(\frac{E(H)}{k_B T_e}\right)$ with $E(H)$ the field-dependent energy barrier to reversal. As suggested by the stretched exponential dynamics, we assume that the magnetization reversal occurs via multiple small independent reversal sites. As the process is dominated by the strong perpendicular magnetic anisotropy, we assume a generic dependence on the field $H$ as $E(H)=E_0\Big(1-\frac{H}{H_a}\Big)^n$ where $H_a$ is the anisotropy field. The model entails approximations on the uniformity of the temperatures $T_e$ and $T_p$, or on the equilibration times of the baths. We discuss these approximations and hypotheses in \cite{supplement}.

\paragraph{Experimental observations --} In this section, we discuss how our measurements corroborate the predictions of the model. In Fig.\ref{fig:scaling}b, the dependence of the switching time $t_{\rm sw}$ on the pulse amplitude $V$ can be well fitted by the Arrhenius law $t_{\rm sw}(H,V)= t_0 \exp\Big(\frac{A(H)}{V^{2/\alpha}}\Big)$ predicted by our model. Typically, we find the so-called attempt times $t_0\simeq0.8-\SI{9}{\nano\second}$, increasing with $H$. Notably, a good agreement is found for Sample B when setting $\alpha = 3.8$, independently determined by measuring the change of longitudinal resistance $R_{xx}$ with temperature $T_e$ or under the application of a DC current (see \cite{supplement,roeper2025}). This rough calibration measurement yields an order of magnitude of the electronic temperature $T_e$ reached as the sample is pulsed. Using a resistance $R_{xx}\simeq \SI{50}{\kilo\ohm}$, one can estimate a substantial increase to a temperature $T_e$ higher than \SI{10}{\kelvin}, largely exceeding the equilibrium condition close to the lattice temperature $T_p\leq\SI{20}{\milli\kelvin}$.
To better confirm the Arrhenius law, we choose a reference field $\mu_0|H_0|=\SI{600}{\milli\tesla}$, and rescale the normalized time $t_{\rm sw}/t_0$ in each data set by the ratio $A(H)/A(H_0)$. In doing so, all data sets fall on a single curve, as shown on Fig.\ref{fig:model}b, further strengthening a scenario based on thermal activation of the switching rate. Additionally, we have fitted the switching time $t_{\rm sw}$ obtained when varying the magnetic field (Fig.\ref{fig:scaling}d). In the limited range of magnetic fields allowed by our measurements, the data is well fitted and further validates the model. The parameters are however extracted with a large uncertainty: $n\simeq 3$ and $H_a\simeq 1 -\SI{1.6}{\tesla}$.

Our analysis therefore reveals that the current-induced switching in our devices probably originates from the exponential reduction of the switching time $t_{\rm sw}(H,V)$, resulting from the elevation of the electronic temperature $T_e$ when submitted to voltage pulses. It rules out, in our devices, spin-transfer torques as the dominant mechanism.

\section{Conclusion}
Our work reveals the dynamics of magnetization switching thanks to the application of short pulses of high voltage amplitude and stroboscopic measurements of the longitudinal and transverse resistances. It elucidates the mechanism of current-induced magnetization switching in ferromagnetic topological insulators in the QAH regime. In particular, we show that the reversal process is governed by thermal activation facilitated by Joule heating. Our observations, in particular with respect to the orientation of the external magnetic field, are incompatible with spin-transfer torques as was reported in \cite{Yuan2024}. 

Future studies may leverage this measurement technique near quantization $R_{xy}= \pm R_K$ to probe the onset of percolative transport through the disordered landscape of magnetic puddles, as pulses are repeatedly applied. Finally, achieving localized heating will enable local magnetization switching, thus opening opportunities for creating and manipulating domain walls in a controlled fashion, for example for the deeper exploration of the interplay of magnetic and electronic degrees of freedom in this material.

The supporting data and codes for this article are available from Zenodo \cite{zenodo_data}.

\section{Acknowledgements}
\begin{acknowledgments}
    We warmly thank A. Rosch for our insightful discussions. This work has been supported by Germany’s Excellence Strategy (Cluster of Excellence Matter and Light for Quantum Computing ML4Q, EXC 2004/1 - 390534769) and the DFG (SFB1238 Control and Dynamics of Quantum Materials, 277146847, projects A04, B07).
\end{acknowledgments}

\bibliography{Literature}

\clearpage
\newpage

\appendix
\renewcommand{\thefigure}{S\arabic{figure}}
\renewcommand{\thetable}{S\arabic{table}}
\setcounter{figure}{0}
\setcounter{table}{0}

{\Large \bf Supplementary Information}

\vspace{2\baselineskip}

\textsc{Sample preparation and measurement setup}

\subsection{Fabrication of the samples}

In this section, we give a very brief description of the different fabrication steps of our devices. As mentioned, the two measured devices are both made from homogeneously V-doped \SI{8}{\nano\meter}-thick BST. The films are grown using molecular beam epitaxy. Directly after the growth, they are capped with a \SI{4}{\nano\meter} layer of Al$_2$O$_3$ to prevent aging of the film. In the first step of the fabrication, the ohmic contacts are patterned on the film using electron beam lithography. This is followed by wet chemical etching of the Al$_2$O$_3$ using Transene Aluminum Etchant type-D and sputtering of \SI{5}{\nano\meter} Pt and \SI{45}{\nano\meter} Au. The Hall- or H-bar-shaped mesas are defined by optical lithography and etched in a Piranha solution. The resulting devices can be seen in Fig.1a of the main text (Sample A) and Fig.\ref{figSup:Hallbar}a (Sample B). The tested devices have a width of \SI{10}{\micro\meter} to \SI{20}{\micro\meter}, while the length ranges from \SI{50}{\micro\meter} to \SI{480}{\micro\meter}.

\subsection{Measurement configuration}

\begin{figure}
    \centering
    \includegraphics[width=\linewidth]{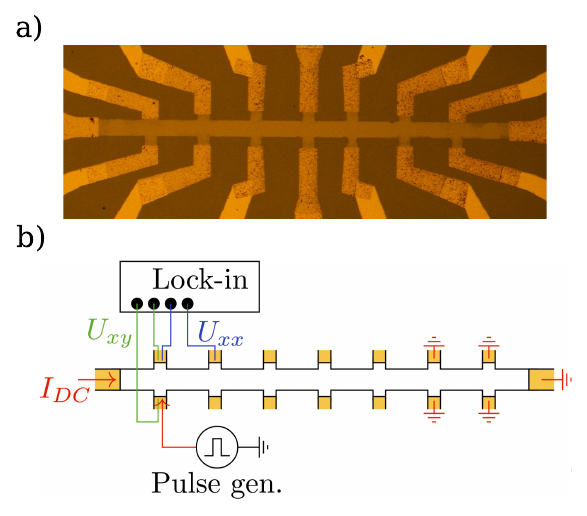}
    \caption{\justifying\textbf{Experimental setup:} a) Microscope image of Sample B, shaped as a long Hall bar. Multiple ohmic contacts are visible on either side of the long mesa structure. b) Schematic of the measurement setup. Similarly to the setup used for Sample A, a pulse generator ({\it Rigol DG5102}) delivers short pulses of duration $\tau$ between \SI{25}{\nano\second} and \SI{100}{\micro\second} applied to the sample via a coaxial line. Lock-in amplifiers measure the longitudinal and transverse (Hall) resistance of the sample in between the pulsed excitations.}
    \label{figSup:Hallbar}
\end{figure}

In Fig.\ref{figSup:Hallbar}a, a microscope image of the second device shown in the main text (Sample B) is shown. It is shaped as a long Hall bar with multiple contacts. This sample has been found to be inhomogeneous in the Hall characterization measurements, and we have therefore focused on the left-hand configuration depicted in Fig.\ref{figSup:Hallbar}b. The pulses are injected in the bottom leftmost contact through a coaxial line while the DC current $I_{DC}$ necessary for the measurement of $R_{xx}$ and $R_{xy}$ is applied on the left contact. Contacts on the opposite side of the Hall-bar are grounded. Overall multiple cool-downs, we have used several types of coaxial lines (thermocoax, semi-rigid cables). Though each type has different bandwidths and different attenuations, it does not affect our analysis where the absolute pulse amplitude is not relevant.

\subsection{Hall measurements}

\begin{figure}
    \centering
    \includegraphics[width=\linewidth]{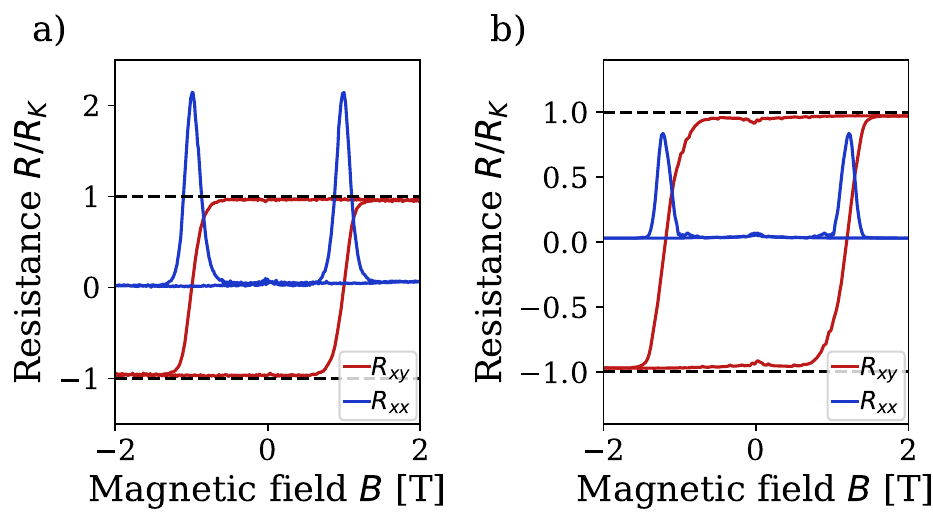}
    \caption{\justifying\textbf{Hall characterization of Samples A and B} a) Measurements of longitudinal and Hall resistances $R_{xx}$ and $R_{xy}$ of sample A, as function of the magnetic field $B$. The expected hysteresis cycle of QAH insulators is observed, and $R_{xy}$ is quantized to $\pm R_K$ within a few percents. The residual longitudinal resistance is typically measured $R_{xx}<\SI{1}{\kilo\ohm}$. b) Similar results for Sample B.}
    \label{figSup:Hall}
\end{figure}

Fig.\ref{figSup:Hall} displays measurements of the longitudinal and Hall resistances $R_{xx}$ and $R_{xy}$ of Samples A and B as function of the magnetic field $B$. Such measurements are performed as basic characterization of all measured devices. They exhibit the expected features of quantum anomalous insulators, i.e. a vanishing longitudinal resistance (here $R_{xx}<\SI{1}{\kilo\ohm}$) and a quantized Hall resistance $R_{xy}= \pm \frac{h}{e^2}$ (here within a few percents). Though the samples do not exhibit a very accurately quantized QAH effect, we argue that this does not significantly hinder our study. Indeed, under applied pulses, the temperature increases to more than \SI{10}{\kelvin}, far from the quantized regime, and where bulk transport is very markedly activated.

\vspace{2\baselineskip}

\textsc{Magnetization dynamics}

\subsection{Dynamics without drive}

In the main text, we have discussed the effect of the pulse duration $\tau$, and have shown that, over a large range of pulse duration (\SI{25}{\nano\second} to \SI{100}{\micro\second}), the only relevant parameter for the magnetization dynamics is the total time spent under applied pulse (denoted $t$).

\begin{figure}
    \centering
    \includegraphics[width=\linewidth]{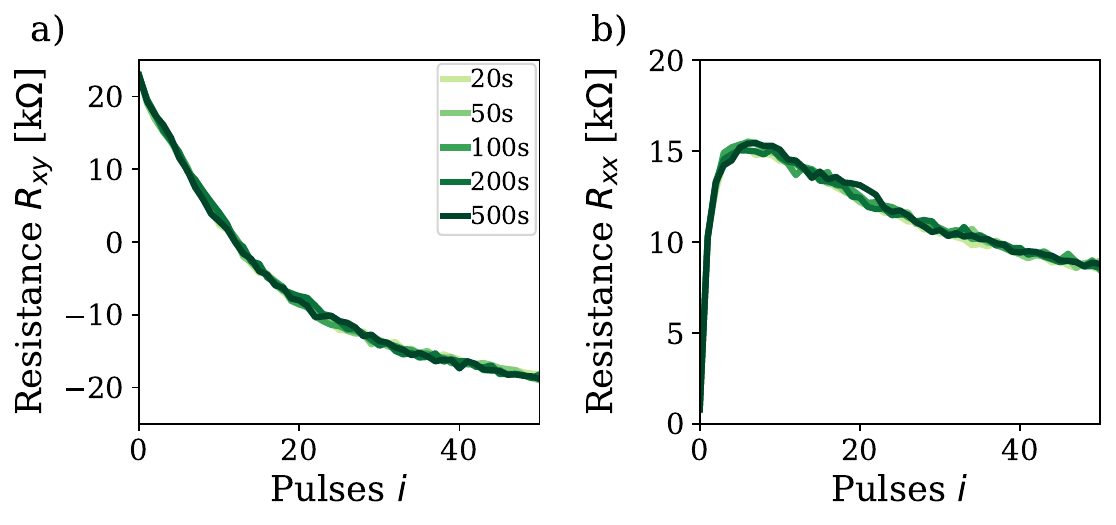}
    \caption{\justifying\textbf{Effect of long time intervals between pulses} Longitudinal and Hall resistances $R_{xx}$ and $R_{xy}$ after $i$ pulses, with a waiting time between pulses varying from \SI{20}{\second} to \SI{500}{\second}. The measurements were taken on Sample A at a magnetic field $\mu_0 |H| = \SI{.5}{\tesla}$, a pulse width $\tau =\SI{50}{\nano\second}$ and a pulse amplitude $V=\SI{5}{\volt}$.}
    \label{figSup:sleep}
\end{figure}

Along the same lines, we illustrate in Fig.\ref{figSup:sleep} that the time interval between two pulses is irrelevant, further showing that the sample dynamics does not evolve significantly between pulses. Fig.\ref{figSup:sleep} shows the resistances $R_{xx}$ and $R_{xy}$ measured after $i$ pulses, for different time intervals between two pulses, from \SI{20}{\second}, showing that this time interval does not modify the measured resistances.

\subsection{Dynamics of the longitudinal resistance}

We observe in some cases that the longitudinal resistance $R_{xx}$ has different initial and final values. We formulate the following possibilities. First, the switching may be incomplete, in particular in regions where the current flow is lower, for example close to the ohmic contacts serving as voltage probes. Second, the geometry may play a role. In the H-bar in particular, the unconventional contact geometry can lead to larger intermixing of the longitudinal and Hall resistance. Finally, the narrower constriction has a finite resistance (even in the magnetized state) and can further complicate the correct estimate of $R_{xx}$ and $R_{xy}$.

\subsection{Fitting procedures}

In the main text, we have analyzed how the switching time $t_{sw}$ behaves as a function of the applied amplitude and magnetic field. Here, we give more details on how we extract $t_{sw}$ as well as on how $t_{sw}(V)$ and $t_{sw}(H)$ were fitted.
\paragraph{Stretched exponentials -- } As described in the main text, we use a stretched exponential to fit the switching dynamics. In order to reduce the number of fit parameters, the resistance $R_0$ is fixed to the value of the first point in each measurement set, i.e. before any pulse at $t=0$. The results for the fit parameter $t_{sw}$ are discussed in detail in the main text. Additionally, we observe that the stretching exponent $\beta$ shows, for all measurement sets, a decreasing behavior with increasing switching times. In sample B, we observe a mismatch between the data and the stretched exponential functions on small time-scales, as can be seen in Fig. \ref{figSup:SamB_switching}. To improve the fit, we add the parameter $R_1$ to the function, which allows us to set the saturating value of $R_{xy}$ to be different from $-R_0$. For the data sets showing fast dynamics, $R_1$ is fixed to be equal to $-R_0$. For the slow switching data sets, $R_1$ is reduced to a value smaller than $R_0$ to allow the fit to converge.

\paragraph{Arrhenius functions -- }The amplitude dependence of the switching time is fitted with the function $t_{\rm sw}(H,V)= t_0 \exp\Big(\frac{A(H)}{V^{2/\alpha}}\Big)$ representing an Arrhenius activation law. Here, we fix $\alpha$ to the values obtained in temperature dependent measurements of $R_{xx}$ which are discussed in the last section of the Supplementary Material. The resulting fit parameters for the different data sets are given in Table \ref{tab:FitAmp} together with the values for $\alpha$.
Concerning the magnetic field dependence, the function $t_{\rm sw}(H,V)= t_0 \exp\Big(B(V)(1-\frac{H}{H_0})^n\Big)$ is deduced from the Arrhenius law. To reduce the number of fit parameters, we fix $n = 3$ for all three data sets, but reasonable fits can be obtained for a range $n=2-4$. The resulting fit parameters are presented in Table \ref{tab:FitMag}.

\setlength{\tabcolsep}{6pt} 
\renewcommand{\arraystretch}{1.2} 

\begin{table}[]
\centering
\begin{tabular}{c|c|c|c|c|c}
          & \textbf{Sam. A}       & \multicolumn{4}{c}{\textbf{Sample B}}  \\ \hline
$B$ [T]        & 0.5  & 0.5 & 0.6 & 0.8 & 0.9 \\ \hline
$t_{0}$ [ns] & 1.7 & 0.8 & 1.2 & 5.9  & 8.7 \\ \hline
A  [\si{\volt^{2/\alpha}}]       & 19.1 & 27.7 & 21.7 & 12.1 & 9.3 \\ \hline
\end{tabular}
\caption{\justifying Fit parameters for the amplitude dependent measurements on Sample A and B at different magnetic fields. For Sample A $\alpha$ was fixed to $\alpha=2.9$ while for Sample B we use $\alpha = 3.8$. The maximal error margins for the fit parameters are: $\Delta t_0 = \pm$ \SI{0.1}{\nano\second} and $\Delta A = \pm~0.02$ .}
\label{tab:FitAmp}
\end{table}

\begin{table}[]
\centering
\begin{tabular}{c|c|c|c}
          & \textbf{Sample A}      & \multicolumn{2}{c}{\textbf{Sample B}}  \\ \hline
Amp. [V]  & 7.5           & 4             & 6             \\ \hline
$t_0$ [ns] & 5.5 & 13.8 & 11.5 \\ \hline
$H_0$ [T]   & 1.07& 1.64 & 1.49 \\ \hline
B         & 22.90 & 32.10 & 28.62 \\ \hline
\end{tabular}
\caption{\justifying Fit parameters for the magnetic field dependent measurements on Sample A and B at different amplitudes. For all fits $n=3$ was fixed. The maximal error margins for the fit parameters are: $\Delta t_0 = \pm$ \SI{0.5}{\nano\second}, $\Delta H_0 = \pm$ \SI{0.005}{\tesla} and $\Delta B = \pm$ 0.05.}
\label{tab:FitMag}
\end{table}

\subsection{Raw data for sample B}

\begin{figure}
    \centering
    \includegraphics[width=\linewidth]{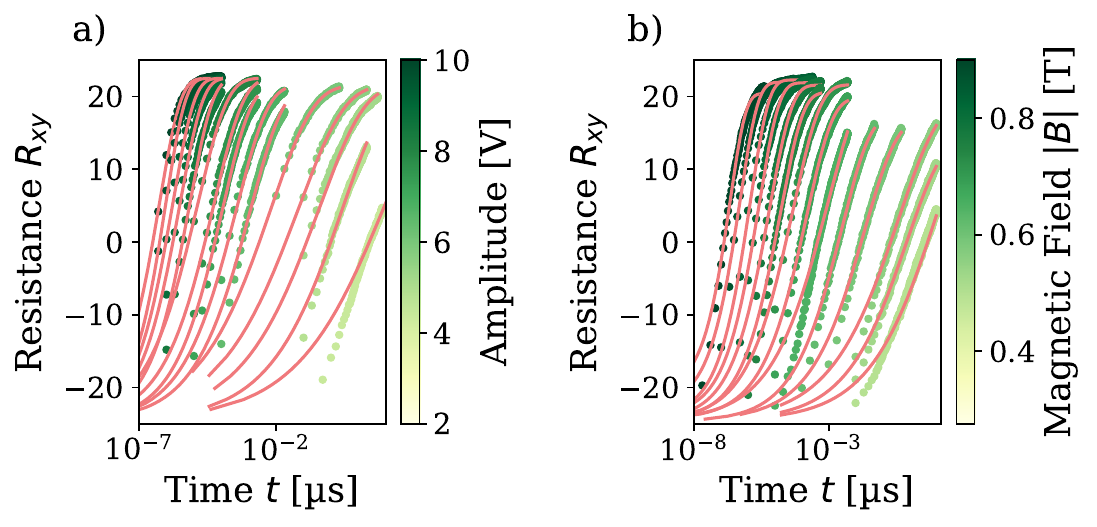}
    \caption{\justifying\textbf{Switching dynamics measured on Sample B} a) Measurements of $R_{xy}$ as function of $t$ for various pulse amplitudes $V$ (shown in different shades of green), taken on Sample B, taken at $\mu_0 |H|=\SI{600}{\milli\tesla}$. b) Measurements of $R_{xy}$ as function of $t$, for various magnetic fields $H$, taken on Sample B, for different magnetic fields at $V=\SI{4}{\volt}$.}
    \label{figSup:SamB_switching}
\end{figure}

In Fig.\ref{figSup:SamB_switching}, we present additional measurements of the dynamics of $R_{xy}$ taken on Sample B, when varying the pulse amplitude $V$ or the magnetic field $H$. The fits to stretched exponentials deviates from data on short time-scales, but reproduce well the long-time tendencies. Despite this shortcoming, these fits provide correct estimates of the switching time $t_{\rm sw}$ that we analyze in the main text.

In addition, we also identified switching in four additional devices, where various issues (large contact resistance, or large longitudinal resistance) precluded a complete study.

\vspace{2\baselineskip}

\textsc{Role of spin-transfer torques}

\subsection{Switching in in-plane magnetic fields}

As mentioned in the main text, we do not observe the reversal of the Hall resistance when only in-plane fields are applied. This contrasts with observations made in Ref.~\onlinecite{Yuan2024} and rules out an important role of spin-transfer torques. This may have several origin. First, we have been so far unable to study the influence of the gate, which is mentioned by the authors as a necessary ingredient. Further studies will explore this aspect. Importantly, Ref.~\onlinecite{Yuan2024} also used Cr-doped BST, with an order of magnitude lower coercive field and they use a tri-layer. Therefore the magnetic and transport of our two layers may strongly differ.

\subsection{Switching under opposite pulse polarity}

To confirm the absence of spin-transfer torques, we studied the reversibility of the magnetization reversal. For spin-orbit torques, the sign of the current determines the applied torque, and therefore allows to switch reversibly the magnetization by inverting the current \cite{Yuan2024}. This work shows that it is not possible in our samples, in the range of explored parameters: after switching, reversing the current direction does not allow to switch back the magnetization to its initial direction.

\begin{figure}
    \centering
    \includegraphics[width=\linewidth]{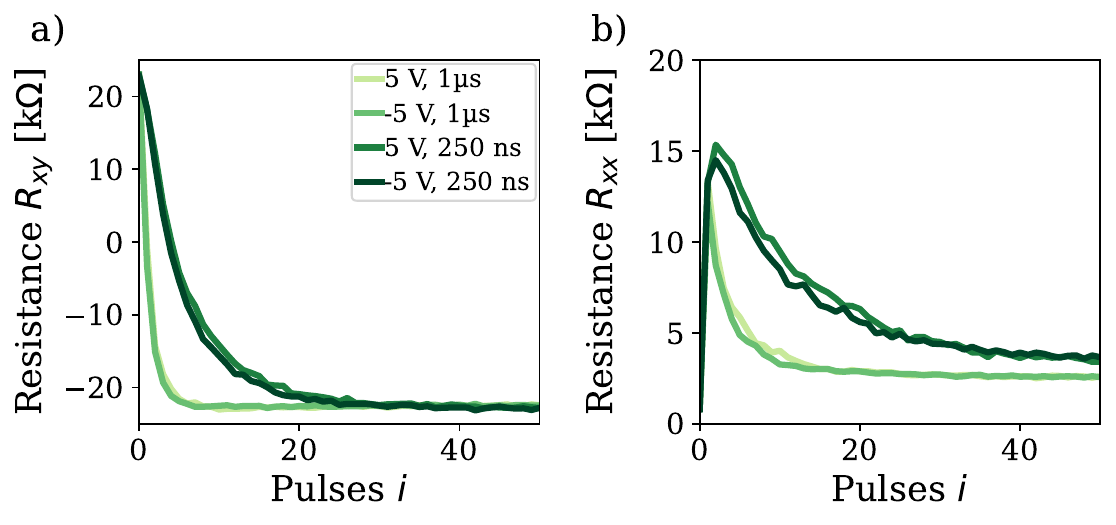}
    \caption{\justifying\textbf{Pulses with opposite polarities} a) and b) Measurements of the Hall resistance $R_{xy}$ (a) and of the longitudinal resistance $R_{xx}$ (b) as function of pulse number $i$, for pulses of opposite polarities $V=\pm \SI{5}{V}$. The data is taken at $\mu_0 |H|=\SI{500}{\milli\tesla}$ on Sample A, for two different pulse durations $\tau = \SI{250}{\nano\second}$ and \SI{1}{\micro\second}.}
    \label{figSup:inversion}
\end{figure}

In Fig.\ref{figSup:inversion}, we show that, for a given field direction (here $\mu_0 |H|=\SI{500}{\milli\tesla}$), the polarity of the voltage pulse does not play a role. In out-of-plane fields (where no spin transfer torques are expected), the longitudinal and Hall resistances  $R_{xx}$ and $R_{xy}$ evolve identically under both amplitudes $V=\pm \SI{5}{V}$. This further consolidates our interpretation, as the dissipated power $P_{\rm diss}$ (hence the temperature $T_e$ and the switching time $t_{\rm sw}$) are identical for both polarities.

\subsection{Switching at opposite fields and cycles}

\begin{figure}
    \centering
    \includegraphics[width=\linewidth]{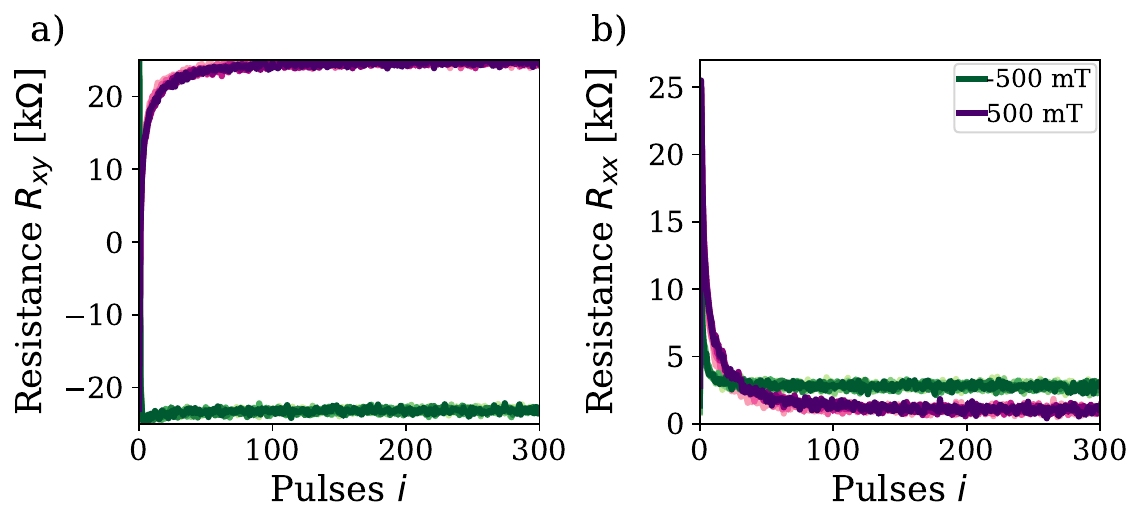}
    \caption{\justifying\textbf{Reversibility of switching} a) and b) Measurements of the Hall resistance $R_{xy}$ (a) and of the longitudinal resistance $R_{xx}$ (b) as function of pulse number $i$ after initial training at $\mu_0 H=\SI{-2}{\tesla}$, following the sequence described in the text. The curves at $\mu_0 H_-$ are shown in shades of green, the ones at $\mu_0 H_+$ in shades of purple. The sequence is more precisely described in the text. The data is taken on Sample A, for $\tau =\SI{1}{\micro\second}$ and $V=\pm \SI{5}{V}$.}
    \label{figSup:cycle}
\end{figure}

We have further explored the reversibility of switching by performing cycles in the magnetic field. Namely, we first fully magnetize the sample uniformly at $\mu_0 H_{\rm init}=\SI{2}{\tesla}$. Then we successfully reversed the magnetization fully at $\mu_0 H_-=\SI{-500}{\milli\tesla}$, then at $\mu_0 H_+=\SI{500}{\milli\tesla}$ (back to its initial direction). We repeated this cycle 5 times in each direction as illustrated in Fig.\ref{figSup:cycle}, without notable differences: though the longitudinal and Hall resistance $R_{xx}$ and $R_{xy}$ evolve differently during forward (at $\mu_0 H_-$) and reverse ($\mu_0 H_+$) switching, no difference is visible from one cycle to the next.

\vspace{2\baselineskip}

\textsc{Heating model}
\subsection{Hypotheses and approximations}

Our model on heating of the electron temperature makes several strong approximations and assumptions.\\

\begin{itemize}
\item {\sl Energy transport --} The energy transport is described by a purely phenomenologically model only with two parameters, $\alpha$ and $\Sigma$.

\item {\sl  Electron and phonon temperatures --} The temperature parameter $T_e$ is defined in the main text as the electron temperature. However, in the section below, where the parameter $\alpha$ is estimated, it is, strictly speaking, the temperature of the conducting electrons as well as that of the phonons which assist the hopping process. 

\item {\sl Homogeneity of the temperatures --} We here assume for simplicity that the temperatures $T_e$ and $T_p$ are uniform over the whole sample. This neglects notably variations of $T_e$ in particular near the corners of the device where the contact resistances dissipate. Differences of the bottom and top surface of the topological insulator layer are not taken into account as their distance (\SI{8}{\nano\meter}) is much shorter than the phonon wavelength in the temperature regime of the experiment.

\end{itemize}

\subsection{Estimating $\alpha$ from DC measurements}

Our model relies on a very simple description of heat transport in our samples, reducing it to two parameters $\alpha$ and $\Sigma$ such that the power transferred to the phonon bath reads: $P_{\rm out}=\Sigma\big(T_e^\alpha-T_p^\alpha)$. Using measurements in the DC regime, we want to estimate of the parameter $\alpha$. 

\begin{figure}
    \centering
    \includegraphics[width=\linewidth]{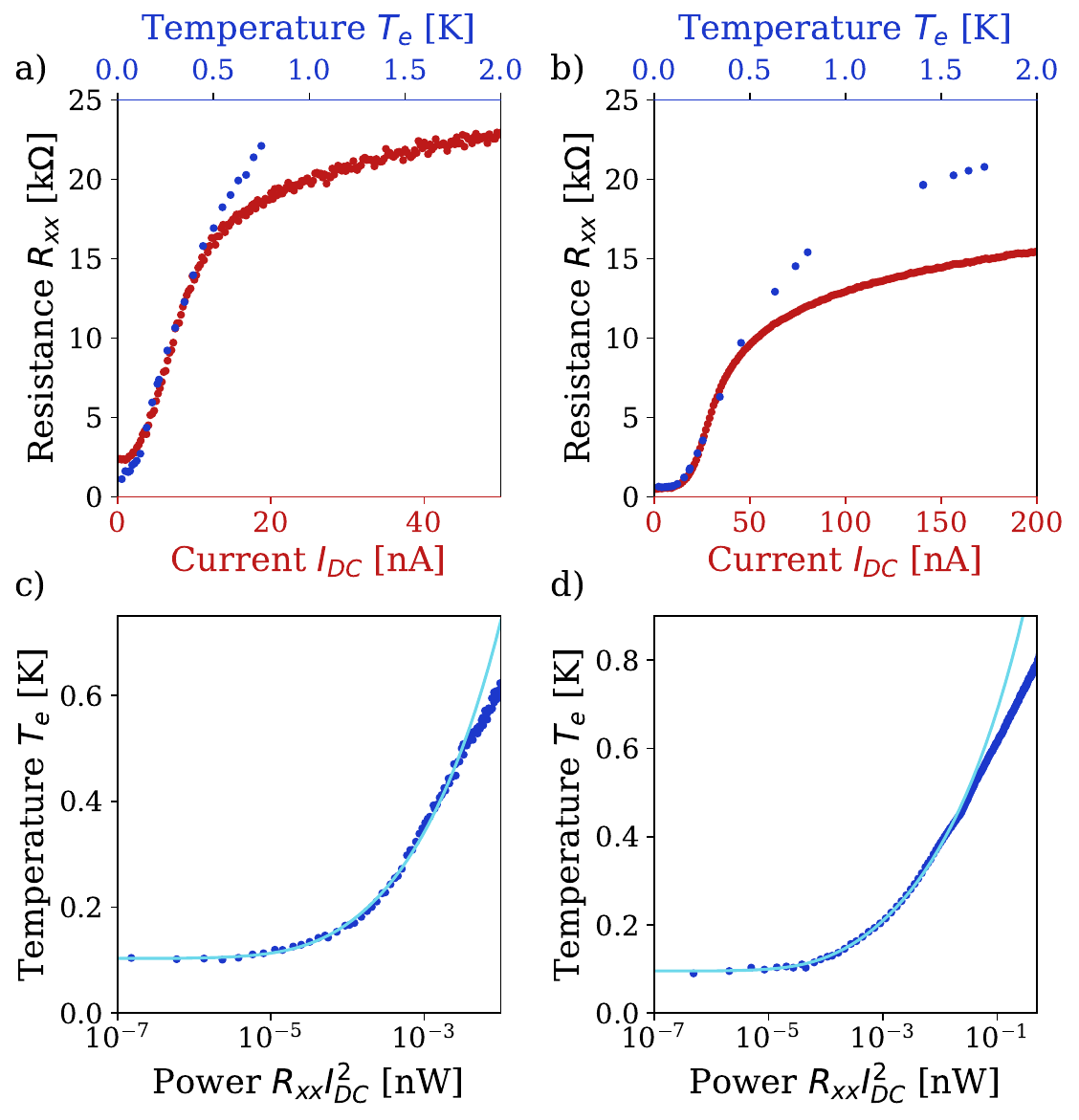}
    \caption{\justifying\textbf{DC measurements to estimate $\alpha$} a) and b) Current and temperature dependent measurements of the longitudinal resistance $R_{xx}$ for Sample A (a) and Sample B (b). c) and d) Extracted temperature $T_e$ in dependence of the applied power for Sample A (c) and Sample B (d). A fit of the data to estimate $\alpha$, as described in the text, is shown in light blue.}
    \label{figSup:DCalpha}
\end{figure}

To this end, we measure the temperature-dependent longitudinal resistance $R_{xx}$ as function of temperature $T_e$ (by heating the refrigerator to known temperatures), and as a function of the input current $I_{DC}$.
Assuming the current $I_{DC}$ leads to a dissipated $P_{\rm out}=R_{xx}I_{DC}^2$, we relate the temperature $T_e$ to the power dissipated $P_{\rm out}$. The curves for $R_{xx}(I_{DC})$ and $R_{xx}(T_e)$ are shown for both samples in Fig. \ref{figSup:DCalpha}a and \ref{figSup:DCalpha}b. Combining the results, we obtain the curves giving the temperature $T_e$ as function of the power $P_{\rm out}$ (plotted in Fig. \ref{figSup:DCalpha}c and \ref{figSup:DCalpha}d), which we fit with the relation $T_e=(T_p^\alpha+P_{\rm out}/\Sigma)^{1/\alpha}$ to estimate $\alpha$. We note that the fits reproduce well the measurements over 4 decades of power. At low temperature, our samples exhibit a residual conductance  (below $T_e\simeq\SI{100}{\milli\kelvin}$). This limits the use of conductance for thermometry in this regime and maps all lower temperatures to $\simeq\SI{100}{\milli\kelvin}$. This limitation, however, does not affect our analysis, as we focus on the high bias where the conductance significantly exceeds this residual value. At higher temperature (above $\simeq\SI{1}{\kelvin}$), other effects than heating (such as thermal activation of bulk carriers, etc) may significantly modify the transport and are not taken into account, leading to further deviations. Therefore, the value of $\alpha$ comes with a large uncertainty (10\%) and should be considered as an estimate. All fit parameters ($\Sigma$, $T_p$, and $\alpha$) are given in Table \ref{tab:FitDC} for the two samples.

\begin{table}[]
\centering
\begin{tabular}{c|c|c}
          & \textbf{Sample A}                       & \textbf{Sample B}                       \\ \hline
$\alpha$  & 2.7 $\pm$ 0.2                  & 3.8 $\pm$ 0.2                  \\ \hline
$T_p$ [K] & 0.1 $\pm$ 0.04                   & 0.1 $\pm$ 0.05                   \\ \hline
$\Sigma [\si{\watt\per\kelvin^\alpha]}$  & (2.4 $\pm$ 0.1)$\cdot$10$^{-11}$ & (4.2 $\pm$ 0.3)$\cdot$10$^{-10}$ \\ \hline
\end{tabular}
\caption{\justifying Fit parameters for the DC calibration curves of Fig.\ref{figSup:DCalpha}c and \ref{figSup:DCalpha}d yielding the parameter $\alpha$.}
\label{tab:FitDC}
\end{table}

\end{document}